\documentclass[pra,twocolumn,showpacs,floatfix]{revtex4}
\usepackage{graphicx,graphics,color,epsfig}
\usepackage{dcolumn}
\usepackage{bm}
\usepackage{amsmath}
\usepackage{amssymb}
\begin{document}
\title{Physical implementation of holonomic quantum computation in decoherence-free subspaces
with trapped ions}
\author{Xin-Ding Zhang}
\affiliation{Department of Materials Science and Engineering,
University of Science and Technology of China, Hefei , China}
\affiliation{School of Physics and Telecommunication Engineering,
South China Normal University, Guangzhou, China}
\author{Qinghua Zhang and Z. D. Wang}
\email{zwang@hkucc.hku.hk} \affiliation{Department of Physics \&
Center of Theoretical and Computational Physics, The University of
Hong Kong, Pokfulam Road, Hong Kong, China}

\begin{abstract}
We propose a feasible scheme to achieve  holonomic quantum
computation in a decoherence-free subspace (DFS) with  trapped ions.
By the application of appropriate  bichromatic laser fields on the
designated ions, we are able to construct  two noncommutable
single-qubit gates and one controlled-phase gate using the holonomic
scenario in the encoded DFS.

\end{abstract}

\date{\today}
\maketitle

Realization of practical quantum computers presents a tremendous
challenge to scientists. One of the most difficulties is caused by
the random errors during gate operations, which mainly roots in
decoherence and stochastic errors in the control process. To
suppress the latter, the holonomic quantum
computation(HQC)~\cite{zanardi} as well as a more general geometric
quantum computation(GQC) scenario~\cite{X.B.Wang,ZW,zhu2003} were
proposed, which are respectively based on the adiabatic non-Abelian
holonomy and the nonadiabatic geometric phase. In the case of HQC,
to realize the wanted gate operation, the Hamiltonian is driven
adiabatically to undergo a designated cycle in the controllable
parameter space, in which degenerate darkstates (with zero
eigenvalues) may be defined. Thus  for an initial state, only the
adiabatic geometric phases can be accumulated in the designated
operation, and the corresponding holonomic quantum gates may be
achieved.  Because of its global geometric feature, it is believed
that the HQC scheme is rather robust against the stochastic errors
occurring in the adiabatic cyclic evolution\cite{zhu2005}. On the
other hand, decoherence is another main obstacle for  quantum
computation. To keep a quantum system away from the main decoherence
source, one should attempt to isolate the system from the
environment that causes mainly the decoherence effect. To deal with
this matter, several scenarios were put forward, including  quantum
error correcting codes~\cite{shor1997} and quantum error avoiding
codes. In particular, when the system-environment interaction has
certain symmetry, a promising scheme based on the
 decoherence-free subspaces(DFS) was proposed
\cite{duanlm,zanardi1997,lidar}. Depending on the type of
interaction symmetry,  one sort of DFS is usually immune to
certain system-bath disturbances.

Very recently, 
combining the HQC with a four-qubit-encoding DFS and based on a kind
of model Hamiltonian, Wu \emph{et al} proposed an interesting
strategy to realize universal quantum computation~\cite{wula}. Such
a strategy was indicated to afford the ability against the
collective decoherence induced by environment and to be rather
robust to certain stochastic errors during operations. Motivated by
the above idea and considering that the four-qubit DFS scenario may
hardly be scalable~\cite{wula}, we here propose a feasible scheme
based on a simple two-qubit-encoding DFS and the adiabatic
non-Abelian holonomy, which is distinctly different from the
four-qubit-encoding scheme. In particular, by the application of
appropriate laser fields, we derive an effective laser-ion
Hamiltonian in a certain kind of trapped ion systems and illustrate
how to physically implement universal honolomic quantum computation
in the present two-qubit-encoding DFS.

\begin{figure}[ht]
\includegraphics[height=1.5cm]{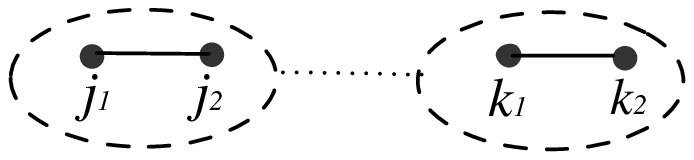}
\caption{A schematic illustration of the encoded sub-systems
(logical qubits) in trapped ions. Each sub-system consists of two
ions enclosed by the dashed circle and the effective interaction
between any two logical qubits is denoted by the dotted line. }
\end{figure}

\begin{figure}[ht]
\includegraphics[height=4cm]{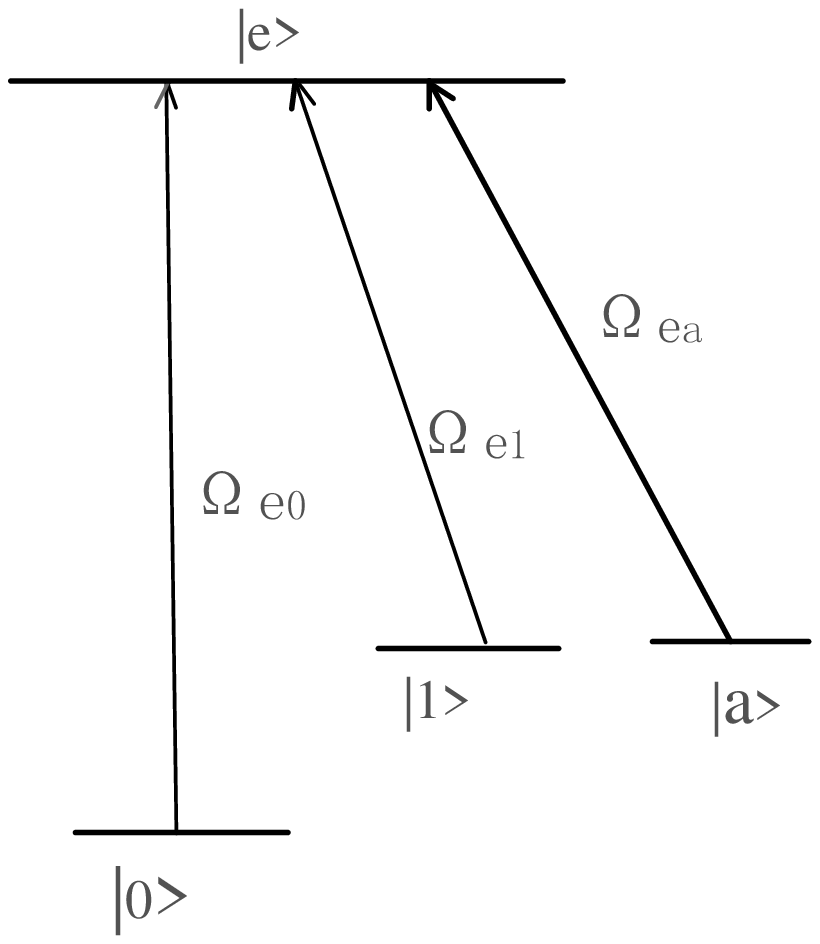}
\caption{Schematic level structure for single ions. $\vert
e\rangle$, $\vert 0\rangle$, and $\vert 1\rangle$ as well as $\vert
a\rangle$ denote the exited state, the lowest ground state, and the
other two stable (metstable) states (with  the last one as an
ancillary state), respectively.  $\vert 0\rangle$ and $\vert
1\rangle$ are the basic units used to encode the logical qubits of
DFS. $\Omega_\alpha$($\alpha=e0,e1,ea$) are the corresponding
Rabi-frequencies with the resonant laser fields being applied on the
ions. }
\end{figure}

To construct the hololomic quantum gates using the decoherence-free
subspaces scheme in a linear Paul trap, we need first to obtain the
wanted effective Hamiltonian for a sub-system consisting of two ions
(e.g., ions $j_1$ and $j_2$ in Fig.1) with the appropriate laser
fields on.
Let us assume that each ion has the four-level energy structure,
as shown in Fig.2~\cite{duan}. The Hamiltonian  may be written
as~\cite{molmer}($\hbar=1$)
\begin{equation}
H=H_{0}+H_{int}
\end{equation}
with
\begin{eqnarray}
H_{0}=\nu(a^{+}a+1/2)+\sum_{i, \alpha}E_{\alpha}\vert
\alpha\rangle_{i}\langle \alpha\vert
\end{eqnarray}
and
\begin{eqnarray}
H_{int}=\sum_{i, \alpha} \Omega^{i}_{e
\alpha}(e^{i[\eta_{i}(a+a^{+})-\omega_i t]}\vert e\rangle_i\langle
\alpha\vert + H.c.),
\end{eqnarray}
where $\alpha\in\{0,1,a\}$ denote the inner states of 
an ion, $\nu$ is the frequency of the center-of-mass vibrational
mode of ions chain, $a^{+}$ and $a$ are the creation and
annihilation operators of the collective motion, $E_{\alpha}$ is the
energy  of the $\vert \alpha\rangle$ state, $\Omega^{i}_{e \alpha}$
is the resonant Rabi frequency between the states $\alpha$ and $e$
of the \emph{ith} ion in the laser field with the frequency
$\omega_i$. The positions of the ions $x_i$ are replaced by ladder
operators $\eta_{i}(a+a^{+})$~\cite{molmer}, where $\eta_{i}$ is the
Lamb-Dicke parameter denoting the ratio of the ion oscillation and
the wave length of the exciting radiation. Since we tune the lasers
close to the the center-of-mass vibrational mode where all ions
vibrate in the same way, the coupling to the vibration is uniform
for all ions, i.e., $\eta_{i}=\eta$ for all $i$.

We now choose the so-called bichromatic laser field~\cite{molmer} to
couple the two ions in the same sub-system (see Fig.2). For the
\emph{$j_{1}$th} ion, we let it be illuminated by two laser pulses,
which are resonant to the transition between $\vert 1\rangle$ and
$\vert e\rangle$, with frequency
$\omega_{j_{1}}=\omega_{e1}+(\nu-\delta_{10})$,
$\omega_{j_{1}}^{'}=\omega_{e1}-(\nu-\delta_{10})$, where
$\omega_{e1}$ denotes the energy difference between the state $\vert
e\rangle$ and $\vert 1\rangle$ and $\delta_{10}$ represents the
additional detuning. In the Lamb-Dicke limit that $\eta^2 (n+1)\ll
1$, where $n$ is the vibrational numbers,  we can approximately
expand the exponential gene $e^{i(\eta (a+a^{+}))}$ to
$1+i\eta(a+a^{+})$. Under the rotating-wave approximation and in the
interaction picture, the laser-ion interaction of the
\emph{$j_{1}$th} takes the form
\begin{equation}
 H_{int}^{j_{1}}=i\eta\Omega^{j_1}_{e1}[ae^{-i\delta_{10} t}+a^{+}e^{i\delta_{10}
t}]\vert e\rangle_{j_1}\langle 1\vert +H.c..
\end{equation}
At the same time, we also apply two laser pulses on the \emph{
$j_{2}$th} ion with $\omega_{j_{2}}=\omega_{e0}+(\nu-\delta_{10})$,
$\omega_{j_{2}}^{'}=\omega_{e0}-(\nu-\delta_{10})$, where
$\omega_{e0}$ denotes the energy difference between the state $\vert
e\rangle$ and $\vert 0\rangle$. Similarly we can have
\begin{equation}
H_{int}^{j_{2}}=i\eta\Omega^{j_2}_{e0}[ae^{-i\delta_{10}
t}+a^{+}e^{i\delta_{10} t}]\vert e\rangle_{j_2}\langle 0\vert +H.c..
\end{equation}

 We here consider only the weak-field regime  $\eta\Omega^i
\ll \delta_{10}$, where only a negligible population is transferred
to the intermediate levels $(n \pm 1)$.  In this case and under the
above laser fields, the transition from $\vert 10n\rangle$ to $\vert
een\rangle$ can be anticipated, and the effective Rabi-frequency may
be evaluated in second-order perturbation theory
\begin{eqnarray}
\Omega_{1}&=&\sum_{m}\frac{\langle een \vert H_{int}\vert m
\rangle\langle m \vert H_{int}\vert
10n\rangle}{E_{m}-E_{10n}-\omega_{m}}\notag\\
&=&-\frac{2\eta^{2}\Omega^{j_1}_{e1}\Omega^{j_2}_{e0}}{\delta_{10}},
\end{eqnarray}
where $H_{int}=H_{int}^{j_1}+H_{int}^{j_2}$, and the intermediate
states $\vert m\rangle$ are $\vert e0(n\pm 1)\rangle$ (with the
corresponding $E_m=E_e+E_0+ (n\pm1)\nu$ and $\omega_m=\omega_{e1}\pm
(\nu-\delta_{10})$) and $\vert 1e(n\pm 1)\rangle$ (with the
corresponding $E_m=E_1+E_e+ (n\pm1)\nu$ and $\omega_m=\omega_{e0}\pm
(\nu-\delta_{10})$) (see Fig.3)~\cite{note1,book}.

\begin{figure}[ht]
\includegraphics[height=4cm]{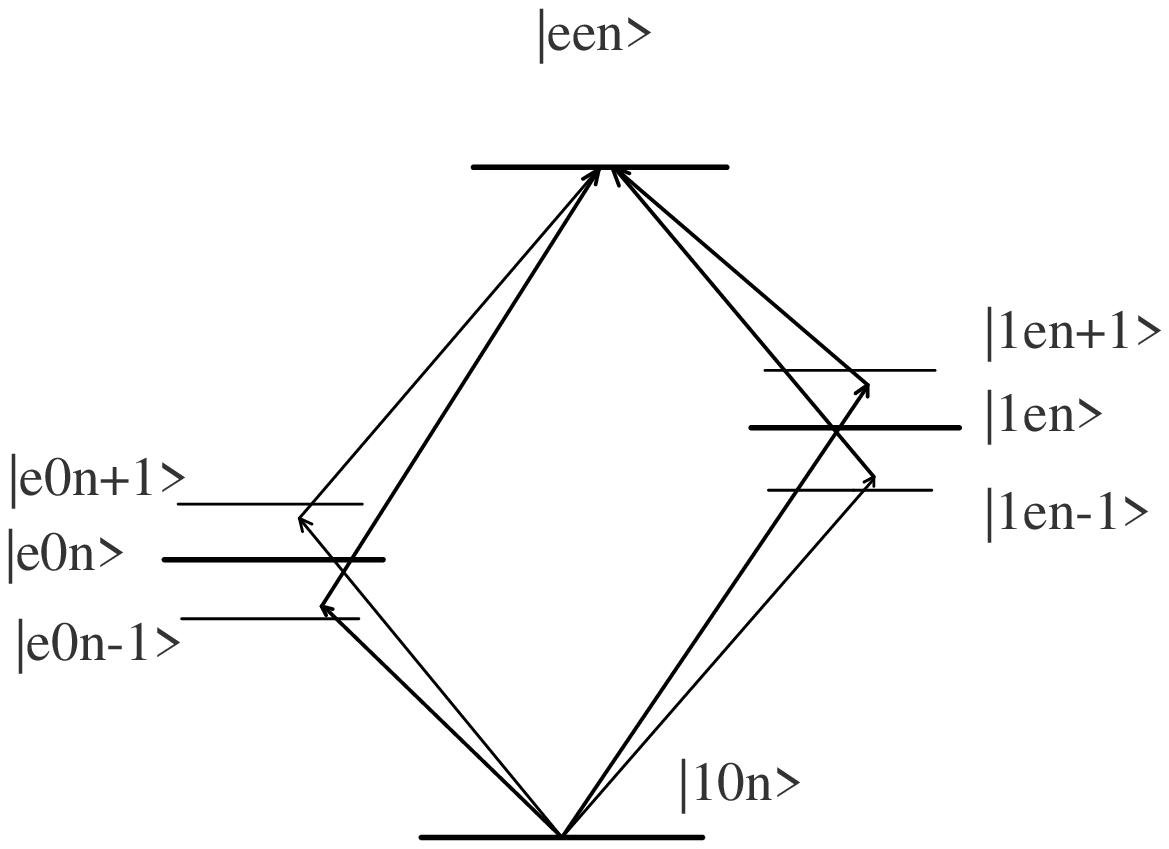}
\caption{Schematic diagram of the resonant transition from $\vert
10n\rangle$ $\rightarrow$ $\vert een\rangle$ for the two ions in a
sub-system. The two ions are illuminated by two different
bichromatic laser fields. Only the transitions involving the
intermediate states with $(n\pm 1$) can occur.}
\end{figure}

Using the same scenario, we can obtain the effective
Rabi-frequencies of the transitions  between ($\vert
01\rangle$,$\vert aa\rangle$) and $\vert ee\rangle$ respectively as
\begin{eqnarray}
\Omega_{0}&=&-\frac{2\eta^{2}\Omega^{j_1}_{e0}\Omega^{j_2}_{e1}}{\delta_{01}}, \notag\\
\Omega_{a}&=&-\frac{2\eta^{2}\Omega_{ea}^{2}}{\delta_{aa}}.
\end{eqnarray}
For simplicity, we may take
$\delta_{10}=\delta_{01}=\delta_{aa}=\delta$ hereafter. Combining
Eqs.(6) and (7) and in the rotating frame,  the effective
Hamiltonian of the sub-system $(j_{1}, j_{2})$ with the lasers on
can approximately be written as
\begin{eqnarray}
H_{eff}&=&[\Omega_{1}\vert ee\rangle\langle 10\vert +\Omega_{0}
\vert ee\rangle\langle 01\vert \notag\\
&&+\Omega_{a}\vert ee\rangle\langle aa\vert ]+H.c.
\end{eqnarray}
in the designated operational subspace spanned by \{$\vert
01\rangle$, $\vert 10\rangle$, $\vert ee\rangle$, $\vert
aa\rangle$\}. At this stage, we consider a smaller subspace
\begin{equation}
C_j :=span\{ \vert 01\rangle, \vert 10\rangle\},
\end{equation}
and encode a pair-bit code to construct a logic qubit in this
subspace: $\vert 0\rangle_{L}=\vert 01\rangle$ and $\vert
1\rangle_{L}=\vert 10\rangle$. If we choose this logic qubit as
computational one, such an encoding constitutes the well known DFS
scheme that is immune from the decoherence induced by the
system-environment interaction in the form of $Z\otimes\textbf{B}$,
where $Z= \sigma_{z}^{j_{1}}+\sigma_{z}^{j_{2}}$ and $\textbf{B}$ is
an random bath operator, simply because for any state
$\vert\psi\rangle\in C_j$, we have $Z\vert\psi\rangle =0$. Also note
that, we can define the corresponding Pauli operators $R_{x}$,
$R_{y}$, $R_{z}$ in  the subspace $C_j$, which can be expressed as
\begin{eqnarray}
R_{x}&=&\frac{1}{2}(\sigma_{x}^{j_{1}}\sigma_{x}^{j_{2}}+\sigma_{y}^{j_{1}}\sigma_{y}^{j_{2}}),\notag\\
R_{y}&=&\frac{1}{2}(\sigma_{y}^{j_{1}}\sigma_{x}^{j_{2}}-\sigma_{x}^{j_{1}}\sigma_{y}^{j_{2}}),\\
R_{z}&=&\frac{1}{2}(\sigma_{z}^{j_{1}}-\sigma_{z}^{j_{2}})\notag,
\end{eqnarray}
where the
$\sigma_{x}^{j},\sigma_{y}^{j},\sigma_{z}^{j}(j=j_{1},j_{2})$ are
the Pauli matrices of the $jth$ qubit.

 We  now address in detail  how
to realize the holonomic computation with a DFS scenario. As is well
known, to achieve a universal set of quantum gates, we need to
construct two noncommutable single qubit operations and a nontrivial
two-qubit gate.
Here we choose the two single logic-qubit gates as
$U_{1}=e^{i\phi_{1}R_{z}}$, $U_{2}=e^{i\phi_{2} R_{y}}$, and the
control phase logic-gate as $U_{3}^{jk}=e^{i\phi_{3}\vert
11\rangle^{jk}_L\langle 11\vert}$,  similar to those discussed in
Refs.\cite{duan,singlebit}.

Denoting also $\vert E\rangle_{L}=\vert ee\rangle$,  $\vert
A\rangle_{L}=\vert aa\rangle$,  the effective Hamiltonian of Eq.(8)
may be rewritten as
\begin{eqnarray}
H_{eff}&=&[\Omega_{1}\vert E\rangle_{L}\langle
1\vert+\Omega_{0}\vert
E\rangle_{L}\langle 0\vert\notag\\
&& + \Omega_{a}\vert E\rangle_{L}\langle A\vert] +H.c..
\end{eqnarray}
To implement the operation $U_{1}$,  we choose $\Omega_{0}=0$,
$\Omega_{1}=\Omega\sin\theta$, and $\Omega_{a}=\Omega\cos\theta
e^{i\varphi}$, where $\Omega$ is the absolute magnitude of the
effective Rabi-frequency for the gate operations, $\theta$ and
$\varphi$ are the control parameters in the parameter space M. In
this case the dark states of the Hamiltonian(11) take the form
\begin{eqnarray}
\vert D_{0}\rangle&=&\vert 0\rangle_L , \notag\\
 \vert D_{1}\rangle&=&\cos\theta\vert
1\rangle_L - \sin\theta e^{-i\varphi}\vert A\rangle_L .
\end{eqnarray}
We then let  $\Omega_{1}$ and $\Omega_{a}$ change adiabatically from
the point \textbf{\emph{O}} ($\theta=0$) along a close path \emph{C}
in the parameter space \emph{M}, i.e., $\theta$ and $\varphi$ take a
cyclic evolution. Thus  for an arbitrary initial state $\vert
\Psi_{0}\rangle$ in the computational subspace $C_j$ , the state
evolves as $\vert \Psi_{0}\rangle \rightarrow U(C) \vert
\Psi_{0}\rangle$ \cite{wilczek, Berry, zhangp}, where
\begin{equation}
U(C)= \exp\oint_{C} A^c
\end{equation}
is the non-Abelian holonomy associated with the path $C$ and
$A^c=\sum_{\mu}A_{\mu}d\lambda_{\mu}$ is the U(2)-valued
connection
expressed as
\begin{eqnarray}
A_{\mu }^{ij}=\left\langle D_{i}\left( \lambda \right) \right\vert \frac{%
\partial }{\partial \lambda _{\mu }}\left\vert D_{j}\left( \lambda \right)
\right\rangle
\end{eqnarray}
where $\lambda_{\mu}$ ($\theta$ or $\varphi$) are the coordinates in
the parameter space. From Eqs.(12)-(14), we can derive the
corresponding connection as
\begin{eqnarray}
A_{\varphi}=-i\frac{1}{2}\sin ^{2}\left( \theta \right) \left(
1-R_{z}\right).
\end{eqnarray}
In this way  the holonomic operation is achieved as
\begin{eqnarray}
U(C) =e^{-i \phi_{1}}e^{i\phi_{1}R_{z}},
\end{eqnarray}
where
\begin{equation}
\phi_1 =\frac{1}{2}\oint_{C} \sin ^{2}\left( \theta \right)
d\varphi.
\end{equation}

Next we deduce  the holonomy $U_{2}$.  We parameterize
$\Omega_{0}=\Omega\sin\theta\cos\varphi$,
$\Omega_{1}=\Omega\sin\theta\sin\varphi$,
$\Omega_{a}=\Omega\cos\theta$. Then the two dark states of
Hamiltonian(11) are given by
\begin{eqnarray}
\vert D_{0}\rangle&=&\cos\theta\cos\varphi\vert
0\rangle_{L}+\cos\theta\sin\varphi\vert 1\rangle_{L}
 -\sin\theta\vert A\rangle_{L},\notag\\
\vert D_{1}\rangle&=&\cos\varphi\vert 1\rangle_{L} - \sin\varphi
\vert 0\rangle_{L}.
\end{eqnarray}
Taking into account another adiabatic cyclic evolution of the
parameters $\theta$ and $\varphi$, we have
\begin{equation}
A_{\varphi}=-i\cos\theta R_{y}\notag .
\end{equation}
Therefore, we obtain
\begin{eqnarray}
U_{2}=e^{i\phi_{2}R_{y}},
\end{eqnarray}
where $\phi_{2}$ is the geometric phase factor determined from the
integral
\begin{eqnarray}
\phi_{2}=-\oint_{C}\cos\theta d\varphi .
\end{eqnarray}

So far, we have illustrated the implementation of two single-qubit
rotations around the y and z axes in the DFS. The combination of
$U_{1}$ and $U_{2}$ allows us to perform any single qubit operation.
To accomplish a set of universal gates, we need to realize one more
two-logic-qubit controlled-phase gate, which is usually more crucial
and important. In our scheme, to realize the two qubit gate
$U_{3}^{jk}$ in the DFS, we propose to couple the $j_{1}th$ and
$k_{1}th$ (see Fig.1) using the bichromatic laser field, and set the
$j_{2}th$ and $k_{2}th$'s energy levels be decoupled. In detail, we
use two resonant laser fields
$\omega_{11}=\omega_{e1}+(\nu-\delta_{11})$ and
$\omega_{11}^{'}=\omega_{e1}-(\nu-\delta_{11})$ to illuminate the
ions $j_{1}$ and $k_{1}$, where $\nu$  still denotes the phonon
frequency of ions.
We denote the effective resonant Rabi-frequency as
$\Omega_{11}^{j_{1}k_{1}}$. The other two pulses
$\omega_{aa}=\omega_{ea}+(\nu-\delta_{aa})$,
$\omega_{aa}^{'}=\omega_{ea}-(\nu-\delta_{aa})$ are used to resonate
the transitions between $\vert a\rangle$ and $\vert e\rangle$ in
both the $j_{1}th$ and $k_{1}th$ ions,
with the corresponding Rabi-frequency as $\Omega_{aa}^{j_{1}k_{1}}$.
As a result, denoting $\vert EE\rangle_{L}=\vert e0\rangle_j \vert
e0\rangle_k$ and $\vert AA\rangle_{L}=\vert a0\rangle_j \vert
a0\rangle_k$,
the effective Hamiltonian of the two sub-systems $j$ and $k$ with
the lasers on
may be written in the coded space as
\begin{eqnarray}
H^{jk}_{eff}=\Omega_{11}\vert EE\rangle_{L}\langle 11\vert +
\Omega_{AA}\vert EE\rangle_{L}\langle AA\vert +H.c.,
\end{eqnarray}
where
\begin{eqnarray}
\Omega_{11}&=&\Omega_{11}^{j_{1}k_{1}}=
-\frac{2\eta^{2}\Omega^{j_{1}}_{e1}\Omega^{k_{1}}_{e1}}{\delta_{11}}, \notag\\
\Omega_{AA}&=&\Omega_{aa}^{j_{1}k_{1}}=-\frac{2\eta^{2}\Omega^{j_{1}}_{ea}\Omega^{k_{1}}_{ea}}{\delta_{aa}}.
\notag
\end{eqnarray}
From the above Eq. (21), similar to the case in the single qubit
operation, we have the two zero-eigenvalue eigenstates (dark states)
as $\vert 00\rangle_{L}$ and $\cos\theta\vert
11\rangle_{L}-\sin\theta e^{i\varphi}\vert AA\rangle_{L}$  under the
parametrization of $|\Omega_{11}|/|\Omega_{AA}|=\tan \theta$ and
$\varphi=\varphi^{AA}-\varphi^{11}$, where $\varphi^{AA}$ and
$\varphi^{11}$ are the phase factors of $\Omega_{AA}$ and
$\Omega_{11}$. When the control parameters $(\theta, \varphi)$ take
a cyclic evolution from $\theta =0$, the component $\vert
11\rangle_{L}$ will accumulate a Berry phase $\phi_{3}$. While at
the same time the other computational bases $\vert 00\rangle_{L}$,
$\vert 01\rangle_{L}$, $\vert 10\rangle_{L}$ are decoupled from the
Hamiltonian(21). This process corresponds a two-qubit
controlled-phase shift gate $U_{3}^{jk}=e^{i\phi_{3}\vert
11\rangle^{jk}_L\langle 11\vert}$ in the DFS.

 We have proposed a
holonomic quantum computation scheme in the DFS with trapped ions.
We have illustrated how to achieve  two noncommutable single
logic-qubit gates and a two-logic-qubit controlled-phase gate using
the non-Abelian holonomy. Note that in our scheme, the computational
bases are all encoded in the subspace immune to the
$\sigma_{z}$-type of decoherence. It is believed that the adiabatic
holonomies combined with the DFS will provide additional virtues
against decoherence and stochastic errors in controlling operations.
It seems feasible  to realize our scheme in physical systems like
trapped ions, though it is very challenging experimentally.

We thank Y. Li and S. L. Zhu for helpful discussions. This work was
supported by the RGC grants of Hong Kong (HKU7114/02P, HKU7045/05P),
the URC fund of HKU, and the NSFC under Grant No.(10429401).



\begin{thebibliography}{99}

\bibitem{zanardi} P. Zanardi and M. Rasetti, Phys. Lett. A \textbf{264}, 94
(1999); J. Pachos, P.Zanardi, and M. Rasetti, Phys. Rev. A
\textbf{61}, 010305(2000).



\bibitem{X.B.Wang} X. B. Wang and M. Keiji, Phys. Rev. Lett.
\textbf{87}, 097901 (2001).

\bibitem{ZW} S. L. Zhu and Z. D. Wang, Phys. Rev. Lett.
\textbf{89}, 097902(2002); Phys. Rev. A \textbf{66}, 042322 (2002);
X. D. Zhang {\it et al.}, Phys. Rev. A \textbf{71}, 014302 (2005).

\bibitem{zhu2003} S. L. Zhu and Z. D. Wang, Phys. Rev. Lett {\bf 91}, 187902
(2003); S. L. Zhu, Z. D. Wang, and P. Zanardi, {\it ibid.} {\bf 94},
100502 (2005).

\bibitem{zhu2005} S. L. Zhu and P. Zanardi, Phys. Rev. A
\textbf{72}, 020301(2005).

\bibitem{shor1997} P. W. Shor, Phys. Rev. A \textbf{52}, 2493(1995)

\bibitem{duanlm} L.M. Duan and G.C. Guo, Phys. Rev. Lett.
\textbf{79}, 1953(1997).

\bibitem{zanardi1997} P. Zanardi and M. Rasetti, Phys. Rev. Lett.
\textbf{79}, 3306(1997).

\bibitem{lidar} D.A. Lidar, I.L. Chuang, and K.B. Whaley, Phys.
Rev. Lett. \textbf{81}, 2594(1998).

\bibitem{wula} L.-A. Wu, P. Zanardi, and D. A. Lidar, Phys. Rev.
Lett. \textbf{95}, 130501(2005).

\bibitem{duan} L. M. Duan, J. I. Cirac, and P. Zoller, Science {\bf 292},
1695 (2001).

\bibitem{molmer} A. Sorensen and K. Molmer, Phys. Rev. A
\textbf{62}, 022311(2000); K. Molmer and A. Sorensen, Phys. Rev.
Lett. \textbf{82}, 1835(1999).

\bibitem{note1} In fact, we may also derive approximately the result in
Eq.(6) in the follwoing way. We rewrite
$H_{int}=\eta(Ae^{i\delta_{10} t}+A^{+}e^{-i\delta_{10} t})$, where
$A=a^{+} [ i \Omega^{j_1}_{e1}(\vert e\rangle_{j_{1}}\langle 1\vert
+ i \Omega^{j_2}_{e0}(\vert
e\rangle_{j_{2}}\langle 0\vert
+H.c.]$. Then in the weak-field regime, the effctive Rabi-frequency
between the addressed transition is approximately given by (see the
Appendix C in Ref.~\cite{book}) $\Omega_1 =\langle een \vert
(\eta^{2}/\delta_{10})[A, A^{+}]\vert
10n\rangle=-2\eta^{2}\Omega^{j_1}_{e1}\Omega^{j_2}_{e0}/\delta_{10}$.

\bibitem{book} C. Gerry and P. Knight, {\it Introductory Quantum Optics},
(Cambridge University Press, 2005).

\bibitem{singlebit} A. Recati, T. Calarco, P. Zanardi, J. I. Cirac and P.
Zoller, Phys. Rev. A {\bf 66}, 032309 (2002).

\bibitem{wilczek} F. Wilczek and A. Zee, Phys. Rev. Lett. {\bf 52}, 2111 (1984).

\bibitem{Berry} M. V. Berry, Proc. R. Soc. London A \textbf{392}, 45(1984).


\bibitem{zhangp} P. Zhang, Z. D. Wang, J. D. Sun, and C. P. Sun, Phys. Rev. A \textbf{71}, 042301
(2005).

\end{thebibliography}
\end{document}